\documentclass[reprint,amssymb, amsmath, aip,cha,twocolumn]{revtex4-1}
\usepackage{graphicx}
\usepackage[caption = false]{subfig}
\usepackage{color}
\usepackage{epsfig}
\usepackage{ifpdf}
\usepackage{url}%
\usepackage{bm}
\usepackage{pdfcomment}
\pdfcommentsetup{draft} 
\pdfcommentsetup{color={1.0 0.0 0.0}, open=true}
\usepackage{cleveref}
\expandafter\ifx\csname package@font\endcsname\relax\else
 \expandafter\expandafter
 \expandafter\usepackage
 \expandafter\expandafter
 \expandafter{\csname package@font\endcsname}%
\fi
\begin{document}
\title{ On the delayed emission from laser produced aluminum plasma under argon environment}%
\author{Garima Arora}%
\email{garimagarora@gmail.com}
\affiliation{Institute For Plasma Research,HBNI, Bhat, Gandhinagar,Gujarat, 382428, India}
\author{Jinto Thomas}
\affiliation{Institute For Plasma Research,HBNI, Bhat, Gandhinagar,Gujarat, 382428, India}
\author{Hem Chandra Joshi}
\affiliation{Institute For Plasma Research,HBNI, Bhat, Gandhinagar,Gujarat, 382428, India}%
\affiliation{Homi Bhabha National Institute, Maharashtra, 400094, India}
\date{\today}
\begin{abstract} 
In this article, we report rather long time emission ($\sim$ 250 $\mu s$) from aluminum neutrals (Al I) in  ns laser produced plasma in the presence of ambient argon. The study is carried out with varying laser power density, background pressure, and the distance from the target. Slow and fast peak components in the emission spectra observed at earlier times are well reported. However, interestingly a very long delayed emission is also observed for the first time which depends on laser power density, distance from the target, ambient gas and pressure. The emission is observed from Al neutrals only. The most likely mechanism of this emission appears to be the excitation and subsequent emission from Al neutrals as a result of energy transfer from metastables of the ambient gas. 
\end{abstract}
\maketitle
\section{Introduction}\label{sec:intro}
Laser Produced Plasma (LPP) has drawn substantial attention of scientific community  due to its numerous applications like lithography \cite{stamm2004extreme,myers2005high,freeman2011enhancements}, nano-particle generation \cite{amoruso2005experimental,rao1995nanoparticle}, cluster formation \cite{tillack2004effect,harilal2013dynamics}, Laser-Induced Breakdown Spectroscopy (LIBS)\cite{musazzi2014laser,singh2020laser,loudyi2009improving,noll2018libs}, radiography\cite{li2006measuring,mackinnon2004proton}, mass spectroscopy\cite{wu2017diagnostic,dyakin1995investigation,okabayashi2011evaluation}, thin-film decomposition\cite{geohegan1994pulsed}, generation of ion sources \cite{barabash1984laser,conzemius1980review}, acceleration\cite{esirkepov1999ion,macchi2013ion}, inertial confinement fusion\cite{betti2016inertial,pfalzner2006introduction}, plasma diagnostics\cite{key1980spectroscopy,glenzer1999thomson,froula2007quenching}, etc. Despite numerous experimental studies and extensive theoretical modeling on LPP's, the plasma plume evolution still presents challenges aspects due to its complexity and transient nature. The lifetime of LPP depends upon the duration of pulse width\cite{le2004influence,margetic2000comparison}, and can be extended by the external factors e.g. background pressure\cite{bulgakov1998gas,skrodzki2016significance,harilal2006ambient,harilal2018optical}.  Investigating plume dynamics in the background medium \cite{sankar2018ion,singh1998effect} has been performed by various groups to understand processes like dragging\cite{sharma2005plume}, thermalization \cite{vaziri2010microscopic}, attenuation/enhancement of optical emissions \cite{kautz2020role}, diffusion\cite{ribiere2009spectroscopic}, recombination\cite{harilal1998influence}, generation of shock waves\cite{gatti1988spherical}, etc. in vacuum where it expands adiabatically. On the other hand the expansion dynamics of plasma plume against a background meduim depends on its atomic mass, pressure, etc. \par
%
Recently, various groups have studied the dynamics of laser-produced plasma in background gas \cite{thomas2018effect,singh2007role,bulgakov1998gas,skrodzki2016significance,harilal2006ambient,harilal2018optical} using multiple diagnostics like Optical Emission Spectroscopy (OES)\cite{aragon2008characterization}, imaging\cite{kuramitsu2011time}, Langmuir Probes\cite{nica2017investigation,kumar2009parametric}, etc. to study the confinement and plume parameters. Among these diagnostics, OES is widely used to study LPP plume evolution due to its simplicity and ability to quantify the parameters like density and temperature due to its high density and the validity of Local Thermodynamic Equilibrium (LTE) condition. Moreover, OES can easily be used to study the evolution of plasma parameters over space and time and provide information about atomic processes e.g. charge exchange process, two and three-body recombinations\cite{rupp1995velocity}, etc.\par 
Previous studies have used imaging and optical time of flight \cite{marine1992analysis,camacho2010time} to investigate the plasma evolution at earlier time scales ($<$10 $\mu s$) where ions and atoms dominate emission. However, emission at a later time scale ($>$20 $\mu s$) has not been studied extensively. Harilal \textit{et. al.}\cite{harilal2011late} reported the fire-work-like emission having Planckian nature from the Graphite target, which is peaked at 20-30 $\mu s$ and existed for 150  $\mu s$ in a vacuum. They have explained this late particle type emission after the onset of plasma is due to material ejection from the sample by the heated gas. Molecular emission has also been reported at such delayed times ($\sim$100 $\mu s$)\cite{de2017laser}. Moreover, the observation of extended emission life time in presences of noble
gases has been correlated  in terms of the thermal conductivity \cite{iida1989atomic} and the collision of metastable states by Kurniawan et. al. \cite{kurniawan1992effect}. Accordingly  the metastable states of inert gases affect the enhanced emission intensity at later time and distance from the sample.\par
Idia \textit{et. al.} \cite{iida1989atomic} reported the emission intensity of non-resonant line of Cu neutrals extend to $\sim$100 $\mu s$ in argon atmosphere. Kurniawan \textit{et. al.}\cite{kurniawan1992effect} used time-resolved spectroscopy to show that two excitation processes can happen in plasma in the presence of noble gases: initially due to the blast wave formation and later due to transfer of energy from metastable state of the noble gases subsequently contributing to even longer emission after the laser irradiation ends.
Diwakar \textit{et. al.} \cite{diwakar2015characterization} also reported the persistence of Cu and Zn species at a later time ($>$50 $\mu s$) from optical time of flight at 10 Torr using ultrafast lasers. They have hypothesized the persistence of species could be due to the metastable state of Ar at high pressure. Recently,  LaHaye \textit{et. al.} \cite{lahaye2021early} combined emission and absorption spectroscopy to characterize the LPP plasma over its full lifetime. Electron excitation and plasma kinetic temperatures are determined using emission spectroscopy at the initial time scale  ($<$5 $\mu s$). Using absorption spectroscopy, they have noticed the presence of aluminum and calcium neutral atoms and their temperature for an extended time scale up to  ($\sim$100-200 $\mu s$). However, emission from neutrals is not reported. \par
A systematic study demonstrating rather long time emission from neutral atoms in laser produced plasma  is naturally important for exploring its origin. Hence in this present work, we report the observation of rather long time emission ($\sim$400 $\mu s$) form aluminum neutrals and its salient features. A Q-switched Nd:YAG laser is used to produce Al LPP. The vacuum chamber, OES system and other instrumentation are described in Sec.(\ref{sec:setup}). In Sec. (\ref{sec:results})  observation of delayed emission is described and a systematic study of this late time emission on laser power density, pressure, and position from the target is discussed. Concluding remarks are added in Sec. (\ref{sec:conc})    \par

\section{Experimental Set-up}\label{sec:setup}
 \begin{figure}[ht]
\includegraphics[scale=0.43]{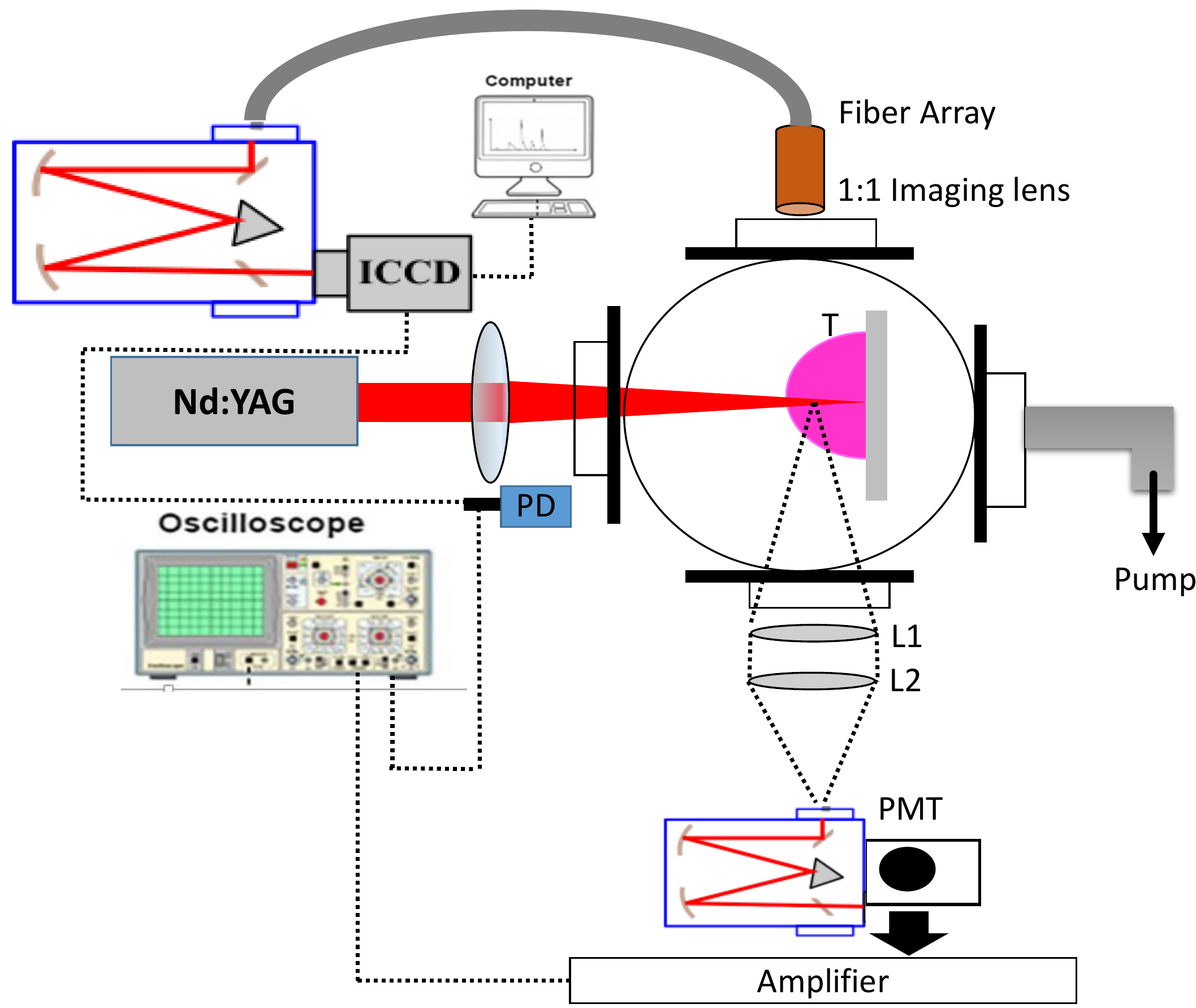}
\caption{\label{fig:fig1} A schematic diagram of experimental set-up. PD is photodiode, ICCD is Intensified Charge Couple Device, PMT is Photomultiplier tube, T is the Aluminum target, L1 and L2 are the lens. }
\end{figure}
Fig.\ref{fig:fig1} shows the schematic diagram of the experimental set-up for studying delayed emission from laser-produced Aluminum (Al) plasma. A Q-switched Nd:YAG laser (wavelength $\lambda=1064~nm$, pulse width $\tau=8~ns$) is used to produce Al plasma. The energy of laser is varied from 100 $mJ$ to 930 $mJ$ and is focused on the sample using a 25 $cm$ plano convex lens. The spot size is fixed at  $\sim 1~mm$ so that laser power density is varied in the range of 2-22 $GW.cm^{-2}$. The energy stability of laser is better than that 3$\%$ .\par 
A cylindrical glass chamber of 100 $mm$ diameter and 600 $mm$ length evacuated to base vacuum of $\sim 10^{-2} ~ mbar$ is used to perform the experiments. A precision gas leak valve is used for filling the chamber with the required gas to the desired background pressure. A calibrated micro pirani gauge is used for measuring the background pressures of respective gases. The Al target is mounted on a translation stage so that a new target position is available for each shot for better reproducibility.\par
 Spectroscopic Time of Flight (STOF) measurements have also been performed using a high-resolution monochromator (Hr460) coupled with a PMT (R943-02 Hamamatsu). The PMT output is connected to a fast digital oscilloscope that records the temporal evolution of optical emission from respective plasma species. 
Another 1 m long Czerny-Turner spectrograph coupled ICCD (Andor DH734) is used to record the emission spectra from the plasma plume. A two lens imaging system with magnification is used for collecting the plasma emission. A fiber array with 8 fibers of 600 micrometer core diameter is used for coupling the image of emitted light into the spectrgraph. The spectrograph with fiber arrays and ICCD enables recording the spectra for different positions simultaneously for a given delay. The spectroscopic and STOF data are averaged for ten observations to reduce the statistical variations in the measurements. In addition to the spectroscopic and STOF data, spectrally resolved fast imaging of the plume is also performed using an ICCD (4 picos Stanford  Computer Optics) and narrow band  interference filters for Al I (396.2 $nm$) line. A Langmuir probe (LP) of diameter 2 $mm$ and length 20 $mm$ is also used in the study. The probe is positively biased to 30 $V$.
\section{Results and Discussion}\label{sec:results}
 \begin{figure}[ht]
\includegraphics[scale=0.45]{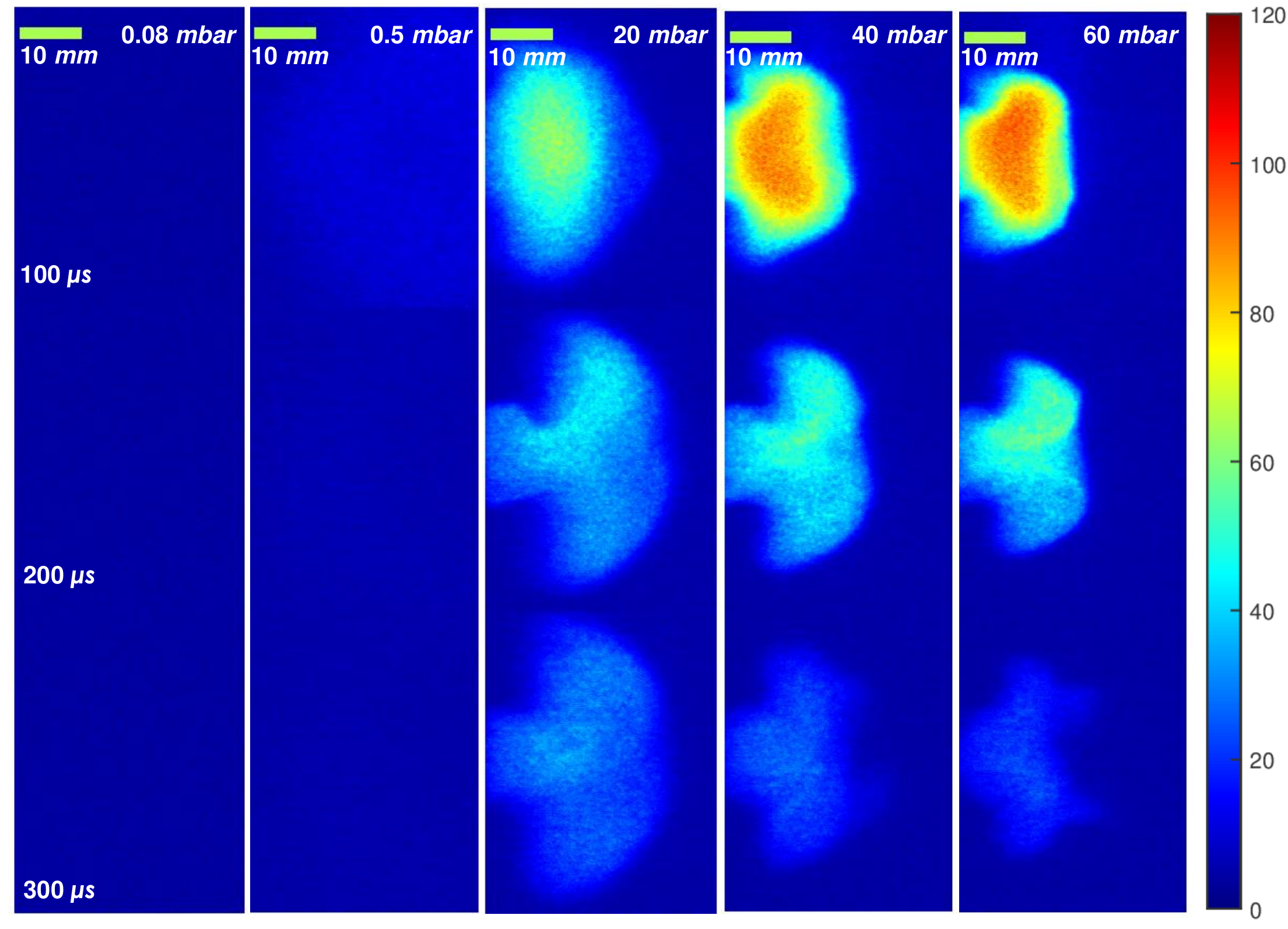}
\caption{\label{fig:fig2}Late time evolution of the images from Al I (396.2 $nm$) line for different ambient pressures of argon recorded using an ICCD and narrow band ( FWHM $\sim$ 0.5 $nm$) interference filter  (396.2 $nm$) for a laser power density of 22 $GW.cm^{-2}$. The marked bar on the top left of the images is 10 $mm$. }
\end{figure}
Fig. \ref{fig:fig2} shows the temporal evolution of aluminium neutral emission recorded using the ICCD and 396.2 $nm$ narrow band interference filter. The integration time for each image is set as 200 $ns$. The images shown are recorded with laser power denisty of 22 $GW.cm^{-2}$ for different ambient pressures of argon. As can be seen from the figure, the delayed emission from aluminum neutral is not evident for lower pressures. However, significant luminosity can be seen for the higher ambient pressures. Morever, at 20 $mbar$ the emission front appears to be moving with rather slow speed (27 $m/s$) which slows down as the pressure increases. The emission lasts longer when the argon pressure is 20 $mbar$ however, at higher pressure the emissivity decreases as the delay increases. This indicates that the confinement of  neutrals with increase in ambient pressures is not as significant as reported for initial times\cite{thomas2018effect}.\par
 \begin{figure}[ht]
\includegraphics[scale=0.32]{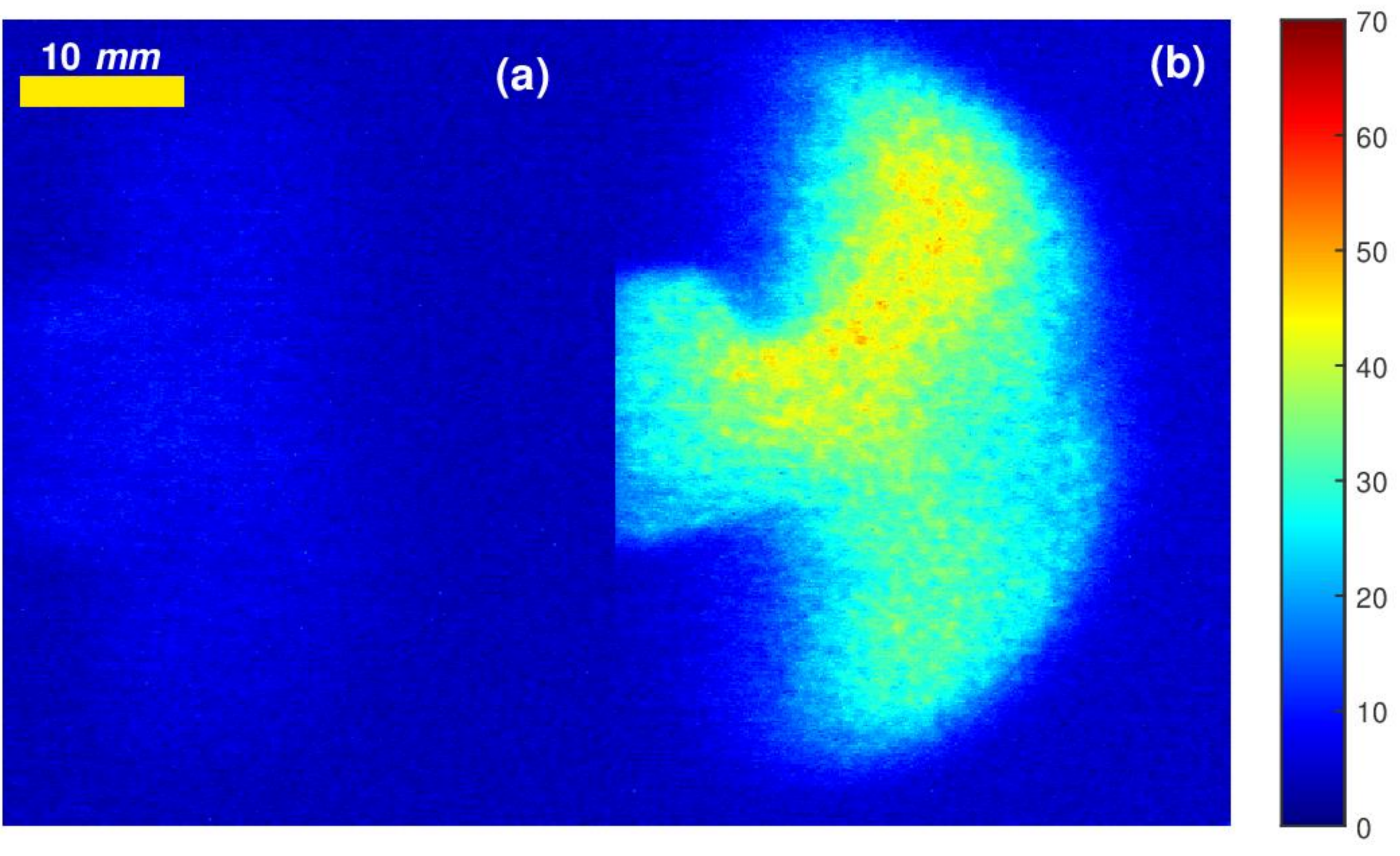}
\caption{\label{fig:fig3} ICCD images of Al I neutral emission recorded at time delay of 200 $\mu s$ for (a) N$_2$ and (b) Ar ambient gas at 20 $mbar$ pressure. The integration time is set to 200 $ns$ and the laser power density is 22 $GW.cm^{-2}$.  }
\end{figure}
Fig. \ref{fig:fig3} shows the snapshot of Al I emission at 200 $\mu s$ recorded from ICCD images for two background gases viz nitrogen and argon for same pressure of 20 $mbar$. One can see clearly that delayed luminosity is present only in case of Ar ambient. \par
 \begin{figure}[ht]
\includegraphics[scale=0.4]{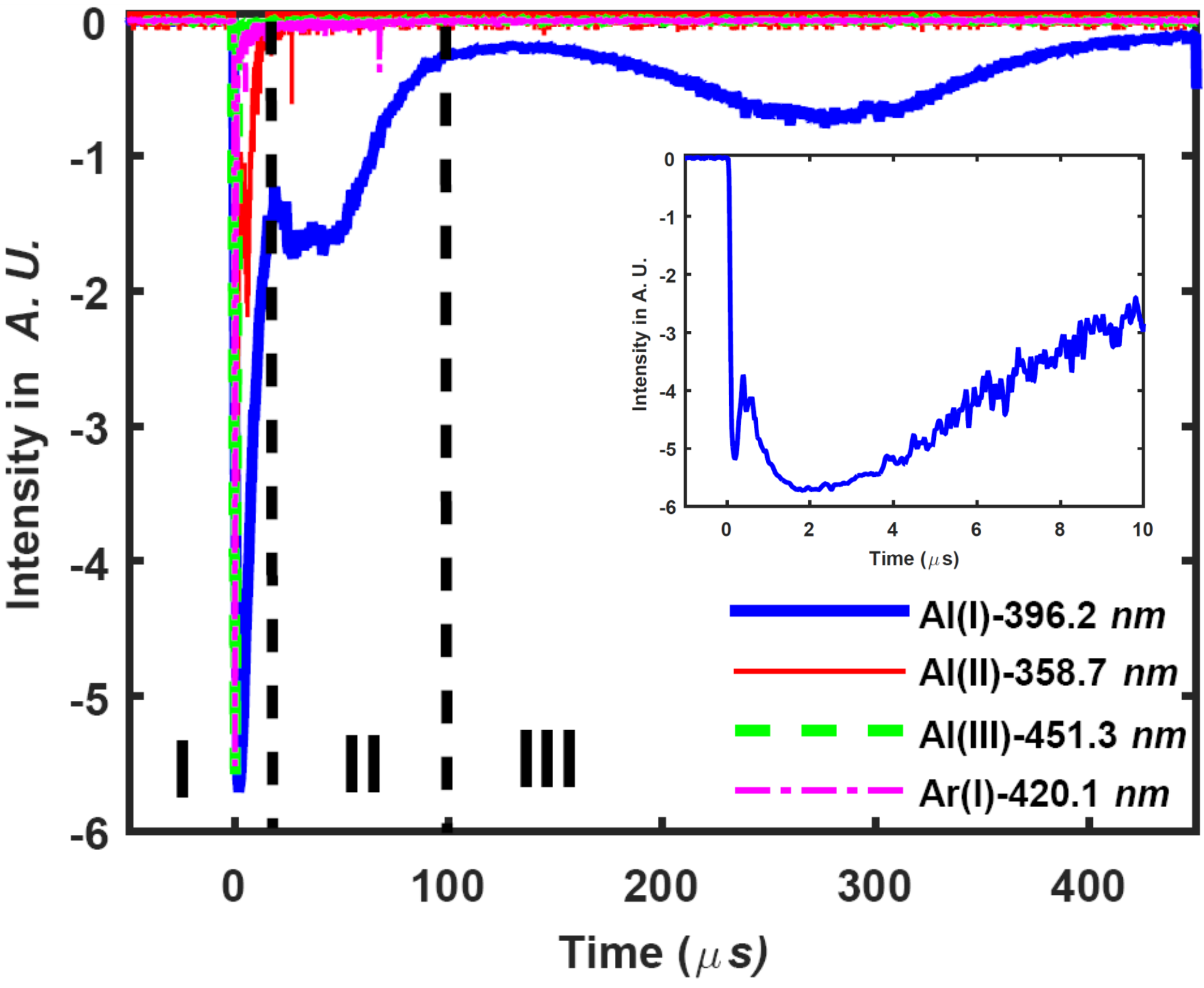}
\caption{\label{fig:fig4}
  STOF profiles of aluminum neutrals (Al I (396.2 $nm$ ($3s^24s\rightarrow~3s^23p$)), aluminum ions (Al II (358.7 $nm$ ($3s4f\rightarrow~3s4d$)), (Al III (451.3 $nm$) ($2p^64d\rightarrow~2p^64p$)) and   argon neutral (Ar I (420.1 $nm$) ($3s^23p^55p\rightarrow~3s^23p^54s$)) recorded at 16 $mm$ from the sample for 22 $GW.cm^{-2}$ laser power density with argon as ambient gas.  }
\end{figure}
Further, STOF studies using high resolution monochromator can provide complete temporal evolution of the respective species for a particular spatial location. Fig. \ref{fig:fig4} shows the STOF spectrum of Al I (396.2 $nm$) (shown in blue solid line) recorded at a distance of 16 $mm$ from the sample for 22 $GW.cm^{-2}$ laser power denisty at 20 $mbar$ argon environemnt. For clarity, we have separated it in three temporal zones as shown in Fig. \ref{fig:fig4}. The first part consists of two peaks with significant intensities. The inset shows the zoomed version of the first region where the first peak is narrow and maximum power comes at $\sim$200 $ns$, and another peak is broad with peak emission intensity at 2 $\mu s$. These neutral emissions are part of the plasma plume produced in the initial time scale after the interaction of laser pulse with the target and propagating in the background medium. The dual peak structure of Al neutrals, as well as ions, at this time scale have been reported earlier \cite{bulgakova2000double} in vacuum as well as ambient pressures in ns laser pulse produced plasma. The fast and slow component structure generation has been explained well and studied in the past \cite{bulgakova2000double,diwakar2015characterization}. The dual peak structure can be attributed to various formation and excitation mechanisms in the ambient gas in a ns laser produced plasma as reported elsewhere \cite{bulgakova2000double,diwakar2015characterization}.\par
 The second region(Region-II) 20 $\mu s$-100 $\mu s$, which peaks at 40 $\mu s$ as can be seen in Fig.\ref{fig:fig4}. Diwakar \textit{et. al.}\cite{diwakar2014expansion} has also observed similar type of delayed peak in emission of Cu and Zn neutrals at 13 $mbar$ of background pressure of Ar using fs laser pulses. They hypothesized that the metastable state of Ar at high pressure can increase the collisions and may lead to an increase in emission intensity of neutral species and increase the persistence of species. We believe a similar type of process is taking place in our case also.\par
However, interestingly the third region (Region-III) shows a broad emission peak of width more than 160 $\mu s$ peaking at around $\sim$ 250 $\mu s$. To the best of our information such a delayed emission as observed in STOF and fast imaging and has not been reported  earlier.\\
It can be noted that STOF evolution of background argon as well as the ionic species of Al do not exhibit such a delayed emission as evident from Fig. \ref{fig:fig4}. It is clear from the figure that the Al neutral line (396.2 $nm$) show peaking around 250 $\mu s$. Delayed emission from Al II and Al III is not observed, as expected because laser-produced plasma has a few microseconds lifetime at similar background pressure and laser power denisty \cite{thomas2020observation} and hence ions are not expected at longer times. It is also worth mentioning that emission at these time scales is not observed from the argon neutrals as well.\par
 \begin{figure}[ht]
\includegraphics[scale=0.38]{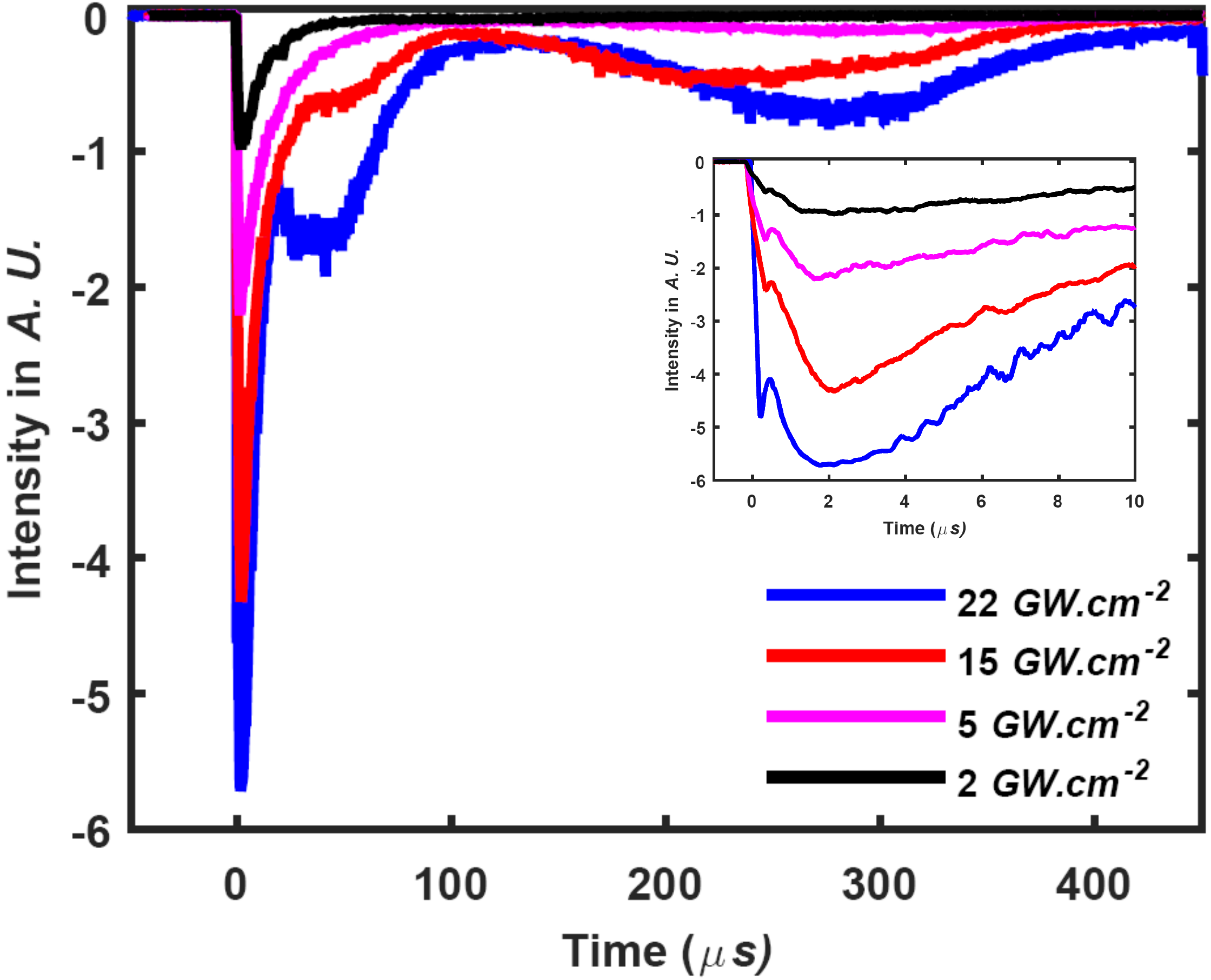}
\caption{\label{fig:fig6} Temporal evolution of neutral emission of Al I (396.2 $nm$) recorded using a fast PMT at 16 $mm$ from the sample for four different laser power densities. The background pressure is set at 20 $mbar$.}
\end{figure}
 \begin{figure*}[ht]
\includegraphics[scale=0.5]{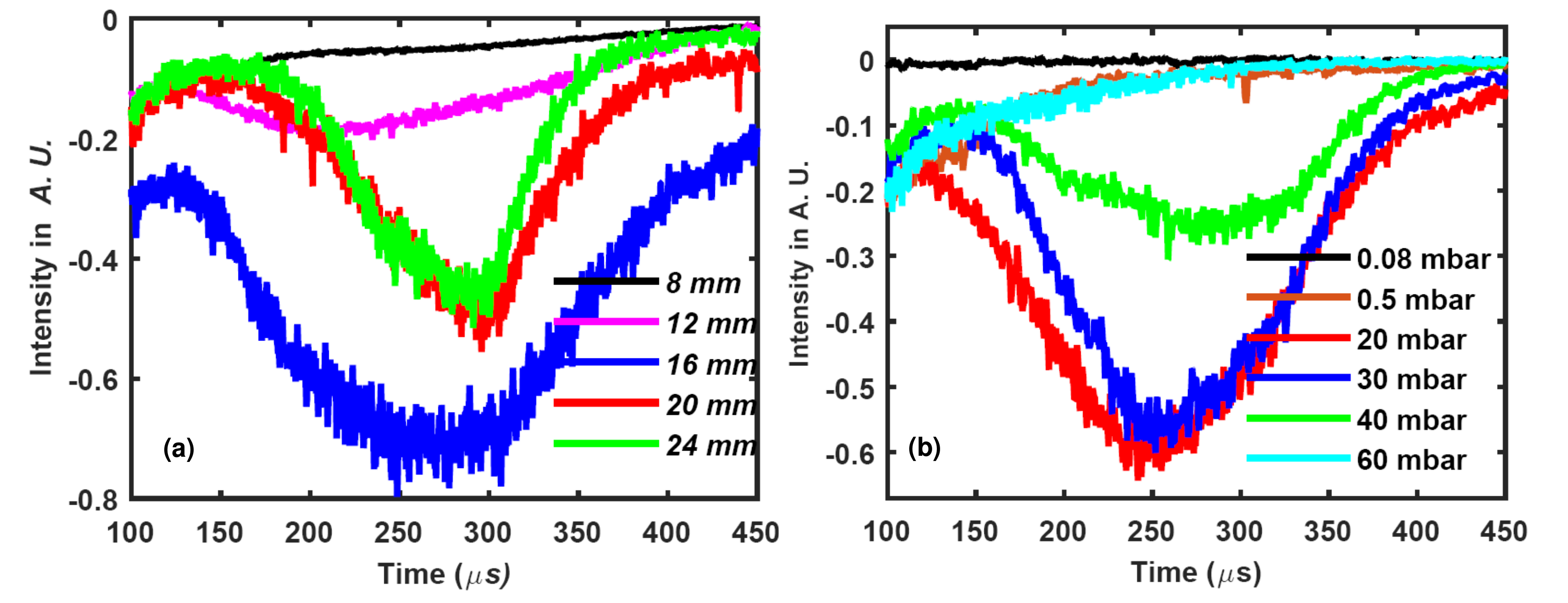}
\caption{\label{fig:fig7} Late time neutral emission of of Al I (396.2 $nm$) is shown for  different (a)  distances from the sample at 20 $mbar$ pressure and (b) Ar background pressures at 16 $mm$ distance. 22 $GW.cm^{-2}$ laser power density is used to produce plasma.}
\end{figure*}
The peaking of Al emission after the termination of plasma appears interesting and a systematic study is carried out to understand the dependence on laser power density, position, type of background gas and background pressure on emission profile of aluminum neutrals (Al I (396.2 $nm$)). Fig. \ref{fig:fig6} shows the effect of laser power density on the emission behavior of Al neutrals at a background pressure of 20 $mbar$. At early stages of plasma plume evolution (shown in inset), the emission intensity increases with an increase in laser power density. As the laser power density increases, the temperature is expected to increase due to the Inverse Bremsstrahlung (IB) process, and hence the emission is enhanced at early times. However, it can be noted that the late-time emission intensity is also enhanced, and the delay of the peak is increased  with laser power density.\par
As can be seen from the figure, at a low laser power density (2  $GW.cm^{-2}$) no emission is observed but as the laser intensities increases, the emission intensity at longer time increases gradually. The delayed emission is prominent for higher laser power density. Here, we would like to mention that fast imaging also show (not shown) that delayed emission is more prominent for higher intensities as observed from STOF profiles.   \par
 Fig.\ref{fig:fig7}(a) shows the spatial evolution of late time emission of Al neutrals at a background pressure of 20 $mbar$ for 22 $GW.cm^{-2}$ laser power density. It can be seen that close to the sample ($\sim$8 $mm$) the emission is rather weak and the peaking behavior is not observed.  However, the emission appears with peaking behavior at 12 $mm$, with maximum at 16 $mm$ , and subsequently decreases. Also the peaking time shows a slight dependence on the position.\par
Fig.\ref{fig:fig7}(b) shows the late emission of neutrals in region-III for  a laser power density of 22 $GW.cm^{-2}$ for different background pressures at 16 $mm$ from the sample. One can see that the emission is not present at lower pressures and appears only at moderate pressures similar to the observations from the fast imaging. The emission appears to peak around 20 $mbar$ and substantially decreases as the pressure increases to 60 $mbar$. It is rather broad, and the peaking time also slightly increases as the background pressure increases, indicating the propagation of the aluminum neutrals is affected by the retardation of the background, which can also be seen from Fig. \ref{fig:fig2}.\par
\par
\begin{figure}[ht]
\includegraphics[scale=0.38]{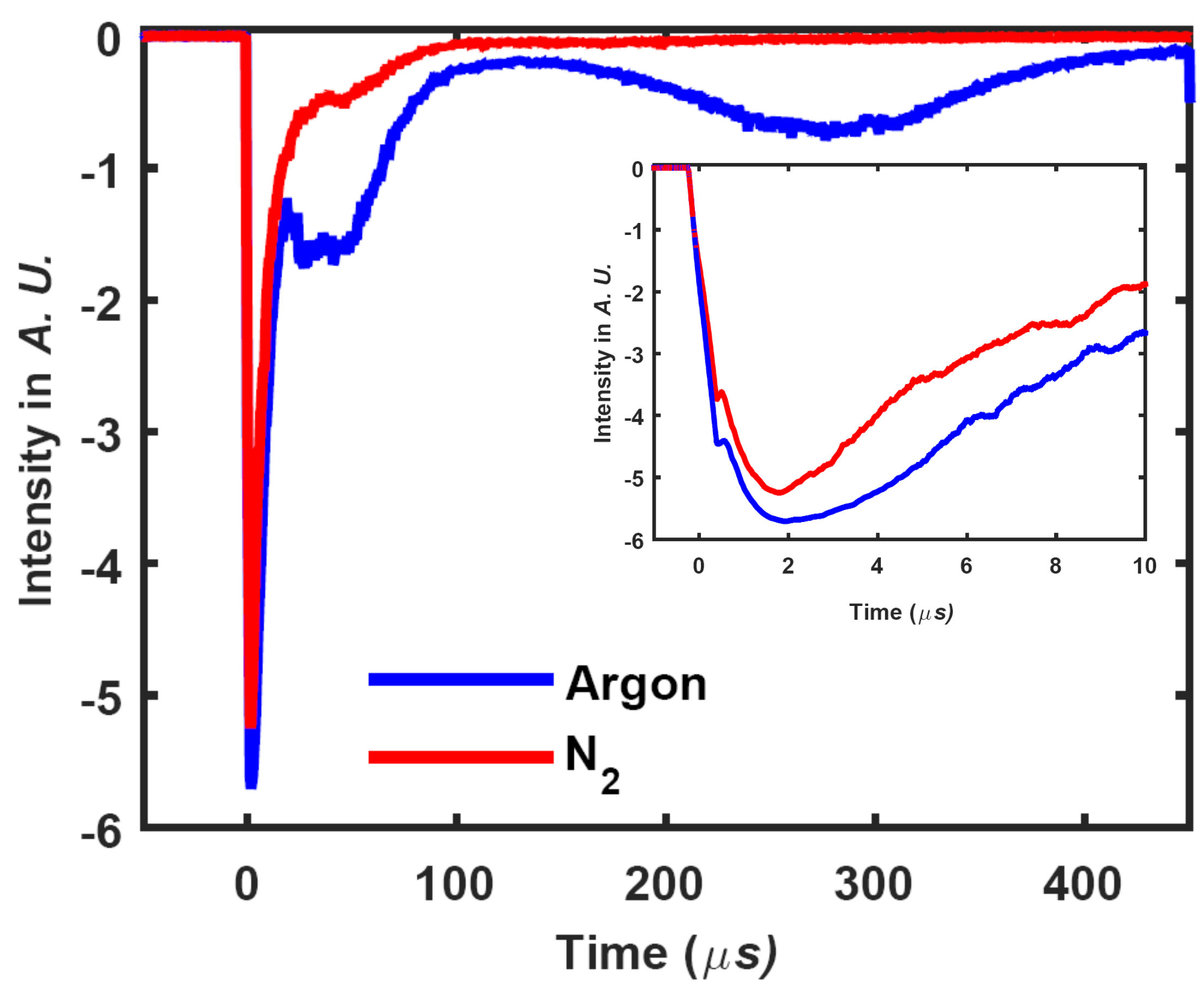}
\caption{\label{fig:fig8} Temporal evolution of Al I (396.2 $nm$) is recorded at 16 $mm$ from the sample at a pressure of 20 $mbar$ for argon and N$_2$ ambient.}
\end{figure}

 In view of this the presence of such a delayed emission from $~~$ Al I appears only in argon environmnet is seen form the ICCD imaging, STOF and spectroscopy data in N$_2$ background is taken to reconfirm the observation of ICCD imaging. The temporal evolution of emission from Al neutrals recorded at 16 $mm$ from the target at 20 $mbar$ background pressure of argon and nitrogen at 22 $GW.cm^{-2}$ laser power denisty is shown in Fig.\ref{fig:fig8}. As can be seen in this figure, the delayed emission from aluminum neutrals is present only for the Ar background medium. As reported in earlier studies of delayed emission up to 20 $\mu s$\cite{diwakar2014expansion}, here also we anticipate that such a delayed emission in the presence of Ar results from the metastable states present in Ar. However, as can be seen in the inset of the figure, the emission dynamics at early stage of the plasma plume remains the same for these background gases. This indicates that background gas does not affect the early dynamics of neutrals as expected from the fact that this emission is due to the shock excitation. However, there is a significant effect of the background gas on the emission dynamics at later times (region-III).\par
\begin{figure}[ht]
\includegraphics[scale=1.1]{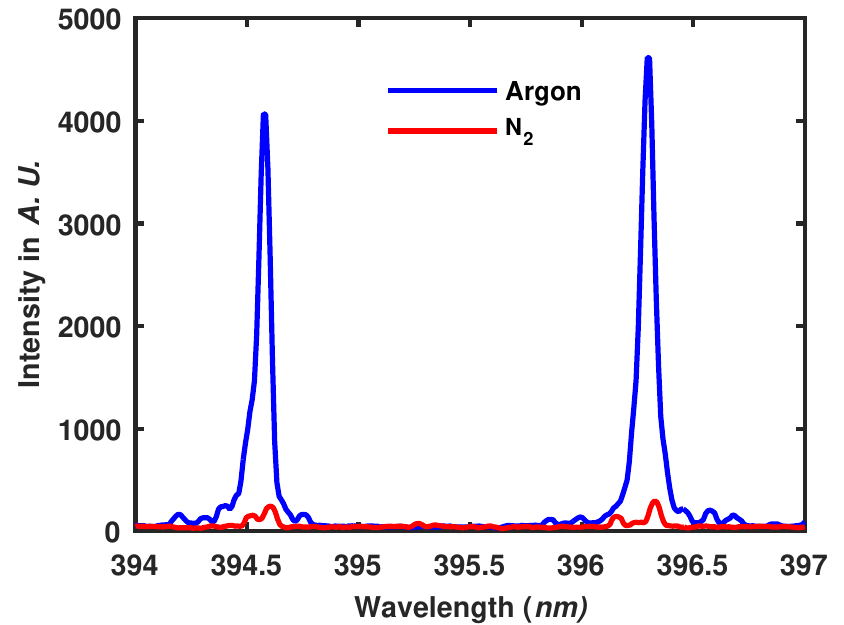}
\caption{\label{fig:fig9} Emission spectra of Al neutrals recorded using a 1 m spectrograph at 16 $mm$ from the sample for argon and N$_2$ ambients. The delay is set at 250 $\mu s$, and the integration time is 20 $\mu s$. }
\end{figure}

To rule out the possibility of molecular emission or incandescent emission, a one meter spectrograph is used to record the line profile of emission and is shown in Fig. \ref{fig:fig9}. Here, the emission spectrum is recorded at a delay of 200 $\mu s$, with a 20 $\mu s$ integration time. The spectra are recorded at 20 $mbar$ pressure and 16 $mm$ from the sample, and the plume is formed by setting 22 $GW.cm^{-2}$ laser power density.  The emission spectra are recorded for Ar as well as N$_2$ ambients and it is also evident from Fig. \ref{fig:fig9} that the emission is present only in the case of Ar ambient. This observation also confirms that the STOF recorded profiles are not contaminated by other emissions. As mentioned, this confirms that the delayed emission is taking place only in the case of argon background.\par
 The possible explanation for this beahviour may be the presensce of metastable states of argon as reported by earlier works\cite{diwakar2014expansion,kurniawan1992effect}. Although, the exact reason for the absence of this emission in N$_2$ is not clear, probably it is due to  the metsatables present in this case may not efficiently populate the excited states of Al neutrals. As the emission from Al neutrals is not evident in case of N$_2$ background for the same pressure, the remote possibility of  the contribution from reflected shock fronts from the walls of chamber can be safely ruled out. Again, as is clear from the following discussion in the fast imaging we do not see any signatures of shock fronts.\par
In order to rule out the possibility of this delayed emission originating from  electrons impact excitation of Al neutrals, measurement from a LP are also carried out. However, it does not give any current after 2 $\mu s$ ruling out the possibility of any significant electron impact excitation of aluminum neutrals for the observed emission. This is also supported by the fact that at longer times electron density is decreased considerably\cite{lahaye2021early}.\par
It can be mentioned here that the long time neutrals emission (around 5 $\mu s$) in the presence of background noble gases, e.g., helium and argon, have been studied earlier also \cite{diwakar2014expansion}. They have postulated a hypothetical model in which two excitation processes in plasma plumes have been taken place. First, primary plasma propagates at high speed and compresses the surrounding gas at higher pressures, forming a blast wave. The second is vaporized atoms which move at slow speed, and meta-stable states of background gases play a role there, which results in more emission at higher pressure. Here also we anticipate that the long lived metastables \citep{diwakar2014expansion} of the ambient gas excite the neutral Al atoms (Ar(metastable)+Al=$Al^*$) evolving slowly after the termination of plasma plume and hence resulting in this late time emission from the aluminum neutral. \par
Further, the absence of delayed emission at lower laser power denisty indicates that the metastable states of Ar are not sufficiently populated owing to the lower energy of electrons in the initial plasma plume. As reported\cite{lichten1957lifetime}, the percentage of population of metastable states and its transition probability depends on the energy of electrons it interacts with. Hence, at lower laser intensities the population density of Ar metastable states is likely to be lower and will be less efficient in transferring energy to the Al neutrals, resulting in a lower emission intensity. As the laser power density increases, the population density of metastable states is likely to increase resulting in more effective collisions with Al neutrals and hence enhanced emission.\par
We have also observed a clear dependence on the position for peaking time and emission intensity (shown in Fig.\ref{fig:fig7}(a)). Again, the change in intensity with distance can be anticipated from the (i) Al neutral density and (ii) Ar metastable population. The absence of emission at very close distances may be due to the fact that at this distance the neutrals are likely to be depleted at this long time scale as the plume itself moves. \par
It is likely that the absence of delayed emission at lower background pressures (shown in Fig.\ref{fig:fig7}(b)) may be due to lack of sufficient density of argon metastables for efficient energy transfer to aluminum neutrals and hence not producing an appreciable emission increase. On the other hand, at higher ambient Ar pressures, quenching of metastables appears to occur which again leads to a decreased energy transfer to neutrals resulting in decreased intensity.\par
 \par
 \begin{figure}[ht]
\includegraphics[scale=0.95]{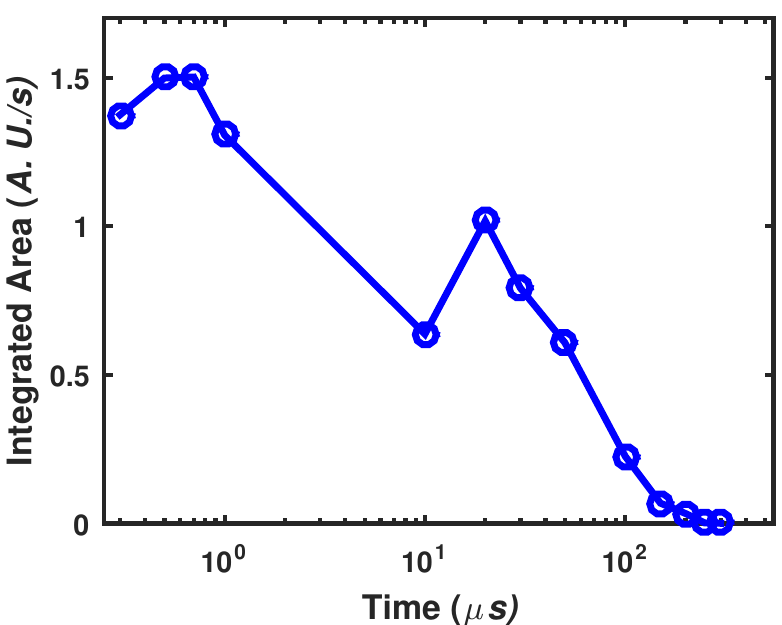}
\caption{\label{fig:fig10} Temporal evolution of Ar I (763.5 $nm$) ($3s^23p^54p\rightarrow~3s^23p^54s$) emission recorded at 16 $mm$ from the sample using high resolution spectrograph at Ar ambient. The gate integration time is 5 $ns$ for initial delays whereas for longer delay it is 1 $\mu s$. The intensities are normalized with the gate integration time. }
\end{figure}
As the most probable reason for the observed phenomena of late emission is the exchange of energy from metastable states of argon with aluminum, we have also recorded the emission of argon neutral for a line which terminates to a metastable state. For this we have taken 763.5 $nm$ line ($3s^23p^54p\rightarrow~3s^23p^54s$) terminating in metastable $^3P_2$ state. Fig. \ref{fig:fig10} shows the temporal evolution of the integrated intensity of the emission of Ar I (763.5 $nm$) line. As can be seen from the figure, the emission from Ar neutrals persists upto 100 $\mu s$ indicating that argon metastables are significantly populated out to this time. This observation further supports the argument regarding possible energy transfer between the metastables of argon with aluminum resulting in such along delayed emission from Al neutrals. Here we would like to mention that energetic metastables of argon can ionize aluminium neutrals by Penning ionization process $Ar^*$(metastable) $+Al~=Al^++Ar+e^-$. However, the observed emission from Al neutrals is unlikely to come from recombination process as electron 
density is too small in this time scale (discussed earlier). Moreover we did not observe emission from Al II states in our experiment.
\section{Conclusions}\label{sec:conc}
 In conclusion, we report a novel interesting feature in the emission of neutrals. The early dynamics show Al neutrals have fast and slow components with the fast peak being narrow and the slow peak broad which is in line with earlier works. However, interestingly the later temporal expansion is made of two parts where the emission intensity shows extrema at 40 $\mu s$ and 250 $\mu s$, which is present only in case of Ar. Moreover, this delayed emission is sensitive to laser power dentistry and position from the target. Further, it is present only at high pressures and is not observed in vacuum. Various metastables of Ar which get populated during early stage of laser plasma interaction could be the source of this excitation of Al neutrals for this delayed emission. In other words these metastable act as excitation energy reservoirs for Al neutrals. These results are also supported by the observed line profiles recorded from the spectrograph and images obtained with ICCD in conjunction with interference filter for Al I line. In brief, this study provides a clear demonstration of rather late time line emission from Al neutrals which get populated due to the metastables of ambient argon. We believe that such a  long time emission from neutral species in a laser produced plasma may open up channels for new applications in this field, particularly in LIBS where in this long time scale, the analysis is expected to suffer from less constraints, particularly the contribution from the emission from other higher ionic states, will insignificant which will help in simplifying the analysis by avoiding the blending of multiple lines. 
%




\providecommand*{\mcitethebibliography}{\thebibliography}
\csname @ifundefined\endcsname{endmcitethebibliography}
{\let\endmcitethebibliography\endthebibliography}{}


\end{document}